\documentclass[a4paper,twocolumn,11pt,accepted=2025-06-26]{quantumarticle}
\pdfoutput=1
\usepackage[colorlinks,bookmarks=false,citecolor=NavyBlue,linkcolor=NavyBlue,urlcolor=blue]{hyperref}

\usepackage{graphicx,bm,amsmath}
\usepackage[usenames,dvipsnames]{xcolor}
\usepackage[bbgreekl]{mathbbol}
\usepackage[normalem]{ulem}
\usepackage{dsfont}
\usepackage{amssymb}
\usepackage{braket}
\usepackage{verbatim}
\usepackage{float}
\usepackage{todonotes}
\usepackage{appendix}

\def\doi{http://dx.doi.org/}
\newcommand{\jdim}{{\rm dim}}
\newcommand{\be}{\begin{equation}}
\newcommand{\ee}{\end{equation}}
\newcommand{\bea}{\begin{eqnarray}}
\newcommand{\eea}{\end{eqnarray}}

\newcommand{\nep}{{\rm e}}
\newcommand{\s}{\mathrm{s}}
\newcommand{\tr}{\mathrm{Tr}}

\definecolor{mypurple}{rgb}{0.49,0.18,0.56}
\definecolor{mygreen}{rgb}{0,0.5,0}
\definecolor{myblue}{rgb}{0,0,0.75}
\definecolor{mymagenta}{cmyk}{0,1,0,0.12}
\definecolor{mygray}{rgb}{0.5,0.5,0.5}

\graphicspath{{./}{./Figures}}

\newcommand{\titleinfo}{Symmetry verification for noisy quantum simulations of non-Abelian lattice gauge theories}

\usepackage[square,numbers,comma,sort&compress]{natbib}
\begin{document}

\title{\titleinfo}

\author{Edoardo~Ballini}
\email{edoardo.ballini@unitn.it}
\orcid{0009-0006-9902-4230}
\affiliation{Pitaevskii BEC Center and Department of Physics, University of Trento, Via Sommarive 14, I-38123 Trento, Italy}
\affiliation{INFN-TIFPA, Trento Institute for Fundamental Physics and Applications, Trento, Italy}

\author{Julius~Mildenberger}
\email{julius.mildenberger@unitn.it}
\orcid{0000-0002-8016-8182}
\affiliation{Pitaevskii BEC Center and Department of Physics, University of Trento, Via Sommarive 14, I-38123 Trento, Italy}
\affiliation{INFN-TIFPA, Trento Institute for Fundamental Physics and Applications, Trento, Italy}

\author{Matteo~M.~Wauters}
\email{matteo.wauters@unitn.it}
\orcid{0000-0003-3261-9425}
\affiliation{Pitaevskii BEC Center and Department of Physics, University of Trento, Via Sommarive 14, I-38123 Trento, Italy}
\affiliation{INFN-TIFPA, Trento Institute for Fundamental Physics and Applications, Trento, Italy}

\author{Philipp~Hauke}
\email{philipp.hauke@unitn.it}
\orcid{0000-0002-0414-1754}
\affiliation{Pitaevskii BEC Center and Department of Physics, University of Trento, Via Sommarive 14, I-38123 Trento, Italy}
\affiliation{INFN-TIFPA, Trento Institute for Fundamental Physics and Applications, Trento, Italy}

\begin{abstract}
Non-Abelian gauge theories underlie our understanding of fundamental forces of modern physics. 
Simulating them on quantum hardware is an outstanding challenge in the rapidly evolving field of quantum simulation. A key prerequisite is the protection of local gauge symmetries against errors that, if unchecked, would lead to unphysical results. 
While an extensive toolkit devoted to identifying, mitigating, and ultimately correcting such errors has been developed for Abelian groups, non-commuting symmetry operators complicate the implementation of similar schemes in non-Abelian theories.
Here, we discuss two techniques for error mitigation through symmetry verification, tailored for non-Abelian lattice gauge theories implemented in noisy qudit hardware: dynamical post-selection (DPS), based on mid-circuit measurements without active feedback, and post-processed symmetry verification (PSV), which combines measurements of correlations between target observables and gauge transformations. 
We illustrate both approaches for the discrete non-Abelian group $D_3$ in 2+1 dimensions, explaining their usefulness for current NISQ devices even in the presence of fast fluctuating noise. 
Our results open new avenues for robust quantum simulation of non-Abelian gauge theories, for further development of error-mitigation techniques, and for measurement-based control methods in qudit platforms.

\end{abstract}

\maketitle

\section{Introduction}\label{sec:intro}
The fast development of quantum technologies over the last decades has stimulated a successful combination of quantum simulations and lattice gauge theories (LGTs)~\cite{dalmonte2016, Preskill_arxiv2018,  banuls2020, Zohar_PhilTransA2021, Aidelsburger_LGT2021, Bauer_PRXQ2023}, paving new ways to advance the understanding of fundamental physics, condensed matter, and quantum information science. 
The intricately constrained dynamics and rich phase diagrams characterizing LGTs are already providing compelling benchmarks for studying and testing the implementation of many-body systems on quantum hardware~\cite{Martinez_Nature2016, Klco_PRA2018,Schweizer_NatPhys2019, Kokail_Nature2019,Klco_PRD2020,Yang_Nature2020, Mil_Science2020, Semeghini_Sci2021, Satzinger_Science2021, mildenberger2022probing, Farrell_PRD2024, meth2023simulating}. 
In order to push this effort further, implementing efficient protocols for non-Abelian gauge groups stands as a major challenge.

Quantum simulations of non-Abelian LGT are a fundamental step toward making quantitative predictions in high-energy physics and are linked to the emergence of universal anyons and topological quantum computation~\cite{Nayak_RMP2008,Shibo_ChinPhysLett2023,Andersen_Nature2023,Iqbal2024}.
However, as quantum simulations are prone to errors and noise from different sources, error mitigation and correction strategies are essential for their successful implementation.
In LGTs, meaningful physical results require the enforcement of an extensive number of local conservation laws. 
For Abelian gauge groups, a comprehensive toolbox has been developed over the years. Examples include adding energy penalty terms to the Hamiltonian to suppress the amplitude of the unphysical sector~\cite{Zohar_PRL2012, Banerjee_PRL2012, Kuno_NJP2015, Dutta_PRA2017, Halimeh_PRL2020, Halimeh_PRXQ2021, Halimeh_NJP2022, mildenberger2022probing}, employing stochastic processes to destroy phase coherence between physical and unphysical spaces~\cite{Stanningel_PRL2014, Lamm_arxiv2020, Kasper_PRD2023, ball2024zeno}, or leveraging local gauge symmetries to select only physical states through a post-selection procedure~\cite{mildenberger2022probing, Nguyen_PRXQ2022, cochran2024stringZ2}.
Although some of these approaches may be expanded to non-Abelian groups \cite{Stanningel_PRL2014, Halimeh_NJP2022, Kasper_PRD2023}, significant gaps persist, in particular regarding post-selection methods and mitigation against incoherent errors. Since the local gauge transformations of non-Abelian theories do not form a commuting set, verifying invariance under one specific local transformation does not necessarily provide any information about whether other transformations are satisfied or not. 
Developing reliable error mitigation schemes remains, therefore, a significant milestone for quantum simulations of LGTs, but also a formidable challenge. 

In this paper, we discuss two modified post-selection algorithms that successfully protect gauge invariance in the digitized dynamics of non-Abelian LGTs against strong random noise. 
We focus on their implementation on qudit devices~~\cite{Wang_FP2020, Wu_PRL2020, Chi_NatComm2022,ringbauer_NP2022, Liu_PRX2023}, which provide the ideal setting to cope with the large local Hilbert space of non-Abelian groups. As a benchmark, we test our protocols on the dihedral group $D_3$, whose small cardinality fits state-of-the-art universal qudit platforms~\cite{ringbauer_NP2022}.  
The first method extends the dynamical post-selection protocol (DPS)~\cite{wauters2024symmetryprotection,schmale2024stabilizing} to non-Abelian groups; it is based on mid-circuit measurements~\cite{Bonet-Monroig_PRA2018, Stryker_PRA2019, schmale2024stabilizing, wauters2024symmetryprotection}, and can be seen as an intermediate step on the way to full-fledged error correction tailored for LGT simulations.
The second is a post-processed symmetry verification (PSV) \cite{Bonet-Monroig_PRA2018, mcclean2017hybrid, mcclean2020decoding, cai2021quantum} extended to LGTs with discrete symmetry groups. 
It is implemented through combining measurements---deriving from the noisy dynamics---of correlations between a desired observable and gauge transformation operators. 
Both methods reliably reconstruct the target dynamics even when the accumulated errors dominate the non-protected dynamics, and even for fast-fluctuating noise without any specific structure. 
These methods are specifically targeted for NISQ devices, since---as with any similar mitigation method---both require an exponential overhead: DPS in the number of acquired shots that satisfy all Gauss' laws and PSV in the number of different operators to be measured or samples of these being drawn. 
Moreover, they complement each other in terms of performance and resource requirements:
on the one side, DPS necessitates mid-circuit measurements but mitigates noise for a longer time. 
On the other side, PSV does not require auxiliary registers and mid-circuit measurements, but is less robust against noise on longer time scales.
Different hardware characteristics and simulation goals can thus favor either one of the two approaches.

The eventual aim of quantum simulation is to enter into \emph{terrae incognitae} beyond the reach of classical computation methods, where validation of the results is far from straightforward~\cite{Leibfried2010,Hauke2012_IOP,Eisert2020,Elben_PRL2020}.  
With fully error-corrected quantum computers still under development, techniques are thus needed to ensure that physics insights extracted from present-day noisy quantum hardware can be trusted.  
The present methods can help fill that gap for quantum simulations of non-Abelian lattice gauge theories. 

The rest of the paper is organized as follows:
To make this work self-contained, in Sec.~\ref{sec:ham} we introduce the formalism needed to implement a quantum simulation of a non-Abelian LGT on qudit devices. In Sec.~\ref{sec:protection}, we explain the general idea of the two techniques for ensuring non-Abelian gauge invariance. Section~\ref{sec:results} specializes the concepts introduced so far to the $D_3$ model and presents our numerical analysis of the two protection schemes.
Finally, we provide an overview of our work and discuss possible developments in Sec.~\ref{sec:conclusion}.
Further numerical results and details are presented in the Appendix.

\section{Hamiltonian real-time evolution for discrete non-Abelian LGTs}\label{sec:ham}

\begin{figure}
    \centering
    \includegraphics[width=0.7\linewidth]{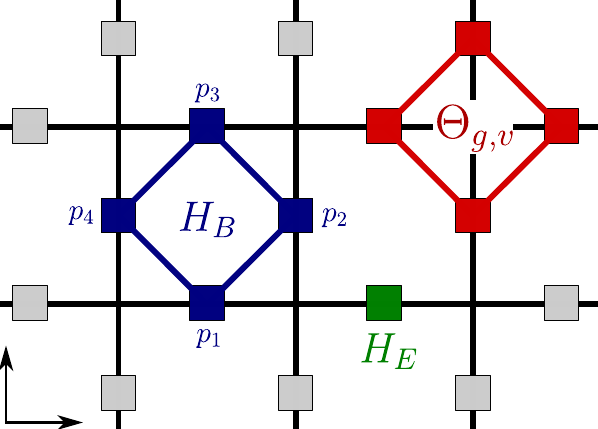}
    \caption{Square lattice with gauge degrees of freedom on the links. Highlighted are a four-body term entering the magnetic Hamiltonian $H_B$, a single-body term of the electric Hamiltonian $H_E$, and a local gauge transformation $\Theta{g,v}$.
    The arrows in the bottom-left corner indicate the directionality of the lattice.}
    \label{fig:lattice}
\end{figure}
\begin{figure*}[ht]
    \centering
    \includegraphics[width=\textwidth]{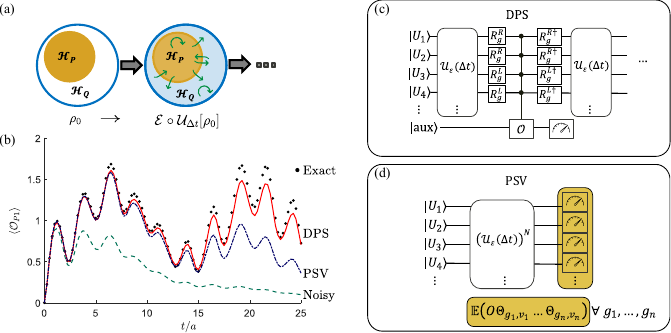}
    \caption{(a) Errors in the quantum simulation of an LGT model couple the physical (gauge-invariant) sector $\mathcal{H}_P$ with the rest of the Hilbert space $\mathcal{H}_Q$. We model errors by a generic noise channel $\mathcal{E}$ added to the desired evolution operator $\mathcal{U}_{\Delta t}$.
    (b) Illustration of the two schemes, DPS (solid line) and PSV (dot-dashed). Both techniques dramatically improve the recovered system dynamics over the simple average over trajectories subject to $\mathcal{E}$ (dashed). DPS is able to follow the exact Trotterized dynamics (bullets) for longer times.
    Ensemble-averages over 5000 trajectories with coupling $1/g^2=0.5$, Trotter step $\Delta t/a =0.25$, and noise strength $\gamma=0.2$. 
    (c-d) Simplified circuit schematics. (c) DPS is based on mid-circuit validations of the Gauss' laws (potentially delaying the measurements and without requiring active feedback). (d) PSV post-processes data acquired from different measurement settings in order to reconstruct gauge-invariant observables. 
    }
    \label{fig:cover_pic}
\end{figure*}
In this work, we are interested in a digital (Trotterized) time evolution of LGT models with a discrete non-Abelian symmetry group.
While many of our concepts are general, we will focus on a pure gauge theory in 2+1 space-time dimensions in the Kogut--Susskind Hamiltonian formulation.
We consider a square lattice with gauge variables residing on its links, see Fig.~\ref{fig:lattice}.
To provide the necessary background, we review the basic formalism for quantum simulations of LGT models with a discrete non-Abelian group $G$, following the notation of Refs.~\cite{ZoharBurrello_PRD2015, Lamm_PRD2019}.

Consider a discrete non-Abelian group with elements $g \in G$ and order $|G| \in \mathbb{N}$. We denote a unitary matrix representation of the group elements as $D^j_{mn}(g)$, where $j$ labels the irreducible representations and $m,n$ are the row and column indices of the matrix.
For finite groups, one has $|G| = \sum_j \jdim(j)^2$, where $\jdim(j)$ is the dimension of the $j$th irreducible representation~\cite{serre1977linear}.
The representation basis $\ket{jmn}$ may be connected to the group element basis $\ket{g}$ following the great orthogonality theorem as 

\begin{equation}\label{eq:GFT}
    \langle g| jmn\rangle = \sqrt{\frac{\jdim(j)}{|G|}}D^j_{mn}(g) \ ,
\end{equation}

which also describes the generalized Fourier transform of the group\footnote{The group Fourier transform is one of the elementary unitary operations needed to implement digital quantum simulations of a lattice gauge theory with discrete groups \cite{Lamm_PRD2019, ZoharBurrello_PRD2015}. To give an example from an Abelian setting, the group Fourier transform for a $\mathbb{Z}_2$ gauge symmetry is the Hadamard gate on qubit Hilbert space.}.
In the following, we will label the computational basis states with the group elements.

The states $\ket{g}$ transform according to the group multiplication. In non-Abelian groups, we must distinguish between left- and right-multiplication operators, which we define in the following way:

\begin{eqnarray}\label{eq:gmul}
    \Theta^L_h\ket{g} & = \ket{hg} \ ,\nonumber  \\
    \Theta^R_h\ket{g} & = \ket{g h^{-1}} \ . 
\end{eqnarray}

These unitary operators enter both the definition of the local gauge symmetries, see Eq.~\eqref{eq:gauss_general} below, and the implementations of the magnetic interaction term, see Appendix \ref{sec:Time_evo_imp}.

Finally, the group connection $U^j$ associates the states $\ket{g}$ with the $j-$th representation of the corresponding group element $U_{mn}^j\ket{g}=D_{mn}^j(g)\ket{g}$.
They transform with the group multiplication operators as
\begin{eqnarray}\label{eq:u_transform}
    \Theta^L_g U^j_{mn} \Theta^{L\dagger}_{g} &= D^j_{mk}(g^{-1})U^j_{kn} \ , \nonumber \\
    \Theta^R_g U^j_{mn} \Theta^{R\dagger}_{g} &= U^j_{mk} D^j_{kn}(g)\ .
\end{eqnarray}
Equation \eqref{eq:u_transform} can be verified by combining Eq.~\eqref{eq:gmul} with the representation of $U^j_{mn}$ in the computational basis $\{ \ket{g} \}$.

With these ingredients, we can describe the pure--gauge Kogut--Susskind Hamiltonian in $2 + 1 $ dimensions~\cite{kogut1979}, which reads
\begin{equation}
    H_{\textrm{PG}}=H_B + H_E\ .
    \label{eq:HPG}
\end{equation} 
The first term of the Hamiltonian is the analog of the energy associated with magnetic fluxes. Its simplest formulation is
\begin{equation}\label{eq:h_v}
    H_B = -\frac{1}{a g^2}\sum_p \Re\left[ {\rm Tr} \left( U^j_{p_1} U^j_{p_2} U^{j\dagger}_{p_3} U^{j\dagger}_{p_4} \right) \right] \ ,
\end{equation}
where $\frac{1}{g^2}$ is the magnetic coupling and $j$ is a faithful irreducible representation. 
We work with natural units $c=\hbar=1$ so that the inverse of the lattice spacing $a$ sets the energy scale. 
The sum runs over all plaquettes in the lattice and $p_1,\dots, p_4$ are the four links of a plaquette taken in anticlockwise order and starting from the lower edge, see Fig.~\ref{fig:lattice}. The directionality of the lattice links determines whether we have to take the operator $U^j_{p_\alpha}$ or its Hermitian conjugate. We set the $+\hat{x}$ and $+\hat{y}$ directions to be positive, consistently with Eq.~\eqref{eq:h_v}.

The second term of the Hamiltonian in Eq.~\eqref{eq:HPG}, $H_E$, represents the action of a local electric field or, in other words, the group Laplacian~\cite{Mariani_PRD2023}. It is diagonal in the representation basis and gauge invariance does not impose a particular choice of its eigenvalues~\cite{ZoharBurrello_PRD2015, Mariani_PRD2023}.
A convenient formulation is derived from the transfer matrix associated with the Wilson action of the magnetic term~\cite{Lamm_PRD2019}, $H_E = - \frac{1}{a}\ln T_E$.
The transfer matrix is defined as 
\begin{equation}\label{eq:Tk}
    \bra{ \mathbf{g}' }T_E \ket{\mathbf{g}} = \prod_l \exp\left\lbrace {\frac{1}{g^2} {\rm Tr}\left[ D^j(g'^{-1}_l) D^j(g_l)\right]  }\right\rbrace \ ,
\end{equation}
where $\ket{\mathbf {g}}$ and $\ket{\mathbf{g}'}$ are two vectors in the group-element basis for the spatial two-dimensional lattice while the product runs over all links $l$.
From Eq.~\eqref{eq:Tk}, it follows that the electric Hamiltonian is the sum of one-body terms acting independently on each link.

Physical observables are invariant under local gauge transformations $\Theta_{g,v}$, which act nontrivially on all four links connected to a given vertex $v$ of the lattice. 
In pure gauge systems, $\Theta_{g,v}$ applies left group multiplications to the outgoing links and right multiplications to the ingoing ones~\cite{ZoharBurrello_PRD2015}
\begin{equation}\label{eq:gauss_general}
    \Theta_{g,v} = \prod_o \Theta^L_{g,o} \prod_i \Theta^R_{g,i} \,.
\end{equation}
When adding dynamical charges, the vertex operators also include the matter transformation under the group action, see Appendix \ref{sec:matter} for more details.

In this work, we consider digital quantum simulations of non-Abelian LGTs, described by the Trotterized time evolution
\be
\label{eq:Trotter_evo}
    e^{-iHt} \simeq \prod_{n=1}^{n_t} e^{-iH_E \Delta t} e^{-i H_B \Delta t} \ ,
\ee
where $n_t=t/\Delta t$ is the number of Trotter steps.
To efficiently encode groups of order $|G|>2$, we specialize our discussion to universal {\em qudit} devices ~\cite{Wang_FP2020, Wu_PRL2020, Chi_NatComm2022,ringbauer_NP2022, Liu_PRX2023}, where each fundamental unit of quantum information possesses more than two levels ($d>2$). This enables the implementation of physically local terms using single qudit gates ~\cite{ Gustafson_PRD2022, GonzalezCuadra_PRL2022}. 
The general framework developed for LGTs on qubit digital quantum simulators \cite{Zohar_PRA2017,Bender_NJoP2018,Lamm_PRD2019, Fromm_EPJQT2023} has already been expanded for qudit quantum hardware~\cite{GonzalezCuadra_PRL2022, Gustafson_PRD2022,Zache_Quantum2023,Popov_PRR2024,calajo2024digital}. Furthermore, seminal proof-of-principle experiments on a truncated $U(1)$ group showcased the promising potential of a compact encoding of the gauge degrees of freedom~\cite{meth2023simulating}. 
In particular, qudit hardware is predicted to significantly reduce the circuit depth and increase the fidelity both for one-body~\cite{Gustafson_PRD2022} and multi-body terms~\cite{GonzalezCuadra_PRL2022} in the Hamiltonian.
 We summarize the essential ingredients and compare qubit and qudit-based implementations in Appendix~\ref{sec:Time_evo_imp}.
For an efficient implementation, we assume the number of levels in the qudit to be at least as large as the group order, $d\ge |G|$.
We discuss the resource advantage for this encoding of the gauge degrees of freedom, in contrast to a standard qubit approach, in  Appendix~\ref{app:d3_details}.

\section{Verification of non-Abelian gauge symmetries}\label{sec:protection}

In present-day devices, quantum simulations are typically affected by errors that can break the symmetries of the target model. For LGTs, this means that the time evolution couples the physical superselection sector with non-physical subspaces. We denote these subspaces with $\mathcal{H_P}$ and $\mathcal{H_Q}$, respectively.
A single Trotter step affected by this {\em gauge non-invariant} time evolution can be modeled as
\begin{equation}\label{eq:gen_Ham}
    \rho(t+\Delta t) =  \mathcal{E}\circ\, \mathcal{U}_{\Delta t} [\rho(t)]\ ,
\end{equation}
where $\mathcal{U}_{\Delta t}$ generates the desired gauge-invariant dynamics, according to Eq.~\eqref{eq:Trotter_evo}, and $\mathcal{E}$ is a generic error channel, typically coupling $\mathcal{H}_P$ and $\mathcal{H}_Q$, as sketched in Fig.~\ref{fig:cover_pic}(a). After sufficiently many such time steps, the errors accumulating during the time evolution induce an effective thermalization and the density matrix will typically approach a steady state.

To preserve the gauge invariance of the system state, we exploit the Gauss' law operators in Eq.~\eqref{eq:gauss_general}. Since $[H_{\textrm{PG}}, \Theta_{g,v}] = 0$ for all $g, \ v$, applying the gauge-symmetry transformations does not affect the target gauge-invariant dynamics on the physical superselection sector
\begin{equation}\label{eq:gauss_cons}
    \Theta_{g,v}\ket{\psi_{\textrm{phys}}} = \ket{\psi_{\textrm{phys}}} \ \forall g,v \ ,\forall \psi_{\rm{phys}} \in \mathcal{H}_P \,.
\end{equation}
Employing symmetries is an established tool for detecting and mitigating errors in the outcomes of quantum computations~\cite{Bonet-Monroig_PRA2018, mcardle2019error-mitigated, mcclean2017hybrid, huggins2021efficient, mcclean2020decoding, cai2021quantum, endo2021hybrid, cai2023quantum}. 
These techniques can become more challenging when coping with LGTs, as one has to verify an extensive number of potentially non-commuting symmetries. 

In the case of an Abelian gauge group, one may simultaneously diagonalize the gauge transformation operators allowing for the concurrent confirmation of all conservation laws.
Runs violating the symmetry at any vertex can then be discarded and only the symmetry-verified runs are used to calculate the quantities of interest. 
This simple direct postselection even permits one to extract relevant observables that qudit-wise commute with the Abelian gauge symmetry, like electric field strength, particle number, or chiral condensate, without the requirement for additional circuit elements. 
Such a scheme has been successfully employed in experiment \cite{Nguyen_PRXQ2022, mildenberger2022probing, cochran2024stringZ2}.
In non-Abelian LGTs, similar direct post-selection methods do not work because transformations associated with different group elements generally do not commute, preventing one from projectively monitoring the full gauge invariance at once. 

In this work, we discuss two complementary techniques to nevertheless significantly improve the quality of the acquired data, as detailed in the rest of this section: (i) mid-circuit measurements can be performed on ancillary qudits that encode the local symmetries, an approach that we call dynamical post-selection (DPS)~\cite{wauters2024symmetryprotection}. 
(ii) One can reconstruct expectation values of gauge-invariant operators through an effective projection on the target superselection sector, performed not by a direct projective measurement but by acquiring and post-processing (gauge-noninvariant) data on a suitable set of observables. The observables that have to be measured correspond to correlators between the target observable and gauge symmetry operators. 
We refer to this approach as post-processed symmetry verification (PSV). 
Panels (c) and (d) of Fig.~\ref{fig:cover_pic} respectively display the circuit schematics of both approaches, while panel (b) showcases their effectiveness in recovering the target dynamics in a quantum simulation of a $D_3$ LGT affected by random noise.  

\subsection{Dynamical post-selection}\label{sec:dps}
\begin{figure}
    \centering
    \includegraphics[width=\linewidth]{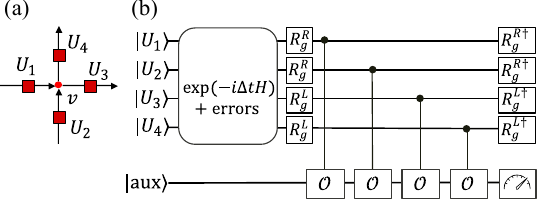}
    \caption{(a) A vertex $v$ and its associated link variables $U_i$. (b) Link variables are encoded in qudits. To remedy the noisy time evolution, we rotate into the basis that diagonalizes the local gauge transformation $\Theta_{g,v}$, whose eigenvalue is then encoded in an auxiliary qudit. The controlled operators $C$-$\mathcal{O}$ depend on the group structure, see the example in Appendix~\ref{app:d3_details}. A local measurement on the auxiliary qudit identifies whether the Gauss' law associated with $\Theta_{g,v}$ is satisfied or not, projecting the wave function onto the corresponding subspace. Finally, we rotate back into the computational basis. After this building block, another Trotter step, another gauge test, or the measurement of desired observables can follow.}
    \label{fig:vertex_circuit}
\end{figure}

Let us consider a general vertex on a lattice as shown in Fig.~\ref{fig:vertex_circuit}(a), where the local gauge transformation associated with a group element $g$ in Eq.~\eqref{eq:gauss_general} reads

\begin{equation}
    \Theta_{g,v} = \Theta_{g,1}^R \otimes \Theta_{g,2}^R \otimes \Theta_{g,3}^L \otimes \Theta_{g,4}^L \,.
\end{equation}

The indices $\{1,\dots,4\}$ label the four links connected to the vertex $v$.
The vertex operator $\Theta_{g,v}$ can be diagonalized by the tensor product of matrices diagonalizing $\Theta^{R(L)}_g$ on each link,

\begin{equation}\label{eq:change_basis}
    R = R_{g,1}^R \otimes R_{g,2}^R \otimes R_{g,3}^L \otimes R_{g,4}^L \ .
\end{equation}

The goal in DPS is to measure the eigenvalue of $\Theta_{g,v}$ without causing a complete wavefunction collapse.
In the spirit of syndrome/stabilizer measurements~\cite{Bonet-Monroig_PRA2018,Stryker_PRA2019,wauters2024symmetryprotection,schmale2024stabilizing},
we introduce an auxiliary qudit to encode the information about the symmetry we want to test.
This auxiliary qudit is initialized in the $\ket{0}$ state and the information about $\Theta_{v,g}$ is encoded through the action of four controlled operators $C$-$\mathcal{O}$, following the basis rotations of Eq.~\eqref{eq:change_basis} to diagonalize $\Theta_{g,v}$, as depicted in Fig.~\ref{fig:vertex_circuit}(b). 
The specific form of the operator $\mathcal{O}$ depends on the group element $g$, but it typically acts on the auxiliary qudit as a controlled permutation of its states.
Then, a local measurement yielding the $\ket{0}$ ($\ket{k\neq 0}$) state on the auxiliary qudit projects the system state onto the subspace that satisfies (does not satisfy) invariance under the target local transformation. Finally, the basis rotation is reversed.
Using qudits ensures all unitaries required for the basis change in Eq.~\eqref{eq:change_basis} can be implemented as single-qudit gates, but generalization to qubit devices is straightforward.

To further understand the result of this procedure, let us consider a situation where the incoherent error channel in Eq.~\eqref{eq:gen_Ham} is described by probabilistically occurring unitaries. Such a scenario includes a broad range of errors, such as uncontrolled gate fluctuations, and will be at the basis of the noise model we use in our numerics in Sec.~\ref{sec:results}. 
Assuming the state at time $t$ satisfied the non-Abelian Gauss' laws, we can approximate a noisy Trotter step of a single random (unitary) trajectory as
\begin{equation} \label{eq:trott_step_DPS}
\begin{split}
    \ket{\psi(t)}_{\textrm{sat}} &\longrightarrow \alpha\ket{\psi(t+\Delta t)}_{\textrm{sat}} + \beta \ket{\psi}_{\textrm{np}} \ \ \\
    &\longrightarrow \alpha\ket{\psi(t + \Delta t)}_{\textrm{sat}}\otimes \ket{0} + \beta \ket{\psi}_{\textrm{np}}\otimes\ket{k \neq 0}  \\
    &\longrightarrow
    \begin{cases}
    \ket{\psi(t+\Delta t)}_{\textrm{sat}} &\textrm{with probability } |\alpha|^2\\
    \ket{\psi}_{\textrm{np}} &\textrm{with probability } |\beta|^2
     \end{cases}\ ,
\end{split}
\end{equation}
where $\ket{\psi}_{\textrm{sat}}$ denotes states that satisfy the local symmetry $\Theta_{g,v}$ that has been monitored and $\ket{\psi}_{\textrm{np}}$ denotes non-physical states that violate it. 
For Abelian groups with a single symmetry generator per vertex, measuring $\ket{0}$ on the auxiliary qudit already implies the output state is locally physical. This is no longer true for non-Abelian groups. 
In that case, the building block of Fig.~\ref{fig:vertex_circuit}(b) can be repeated to check for the other symmetry transformations acting at the same vertex. 
In principle, to obtain a full certification, the gauge transformations at all vertices should be measured. In our numerics below, we cyclically choose only a single transformation at a single vertex in each Trotter step, for which we already find excellent error-mitigation performance.   
Finally, at the end of the procedure, we can measure observables of interest or continue the time evolution.

\subsection{Post-processed symmetry verification}\label{sec:psv}
As pointed out at the beginning of this section, direct post-selection is not accessible for non-Abelian symmetries as the gauge transformations do not all commute pairwise. 
In this section, we outline an alternative approach that leverages the concept of verifying symmetries in post-processing \cite{Bonet-Monroig_PRA2018, mcclean2017hybrid, mcclean2020decoding, cai2021quantum}.
It allows for the verification of gauge invariance without a need for additional circuit elements beyond rotations into a suitable basis at the final step, and is based on acquiring data on an increased set of suitable observables from the error-prone gauge-violating dynamics, from which the gauge-invariant part of the desired observable can be reconstructed. 
Importantly, this approach is valid for any symmetry, Abelian and non-Abelian as well as local and global. 

Generally, the projection of a noisy outcome $\rho$ of a simulation onto the symmetry-preserving subspace, defined by the projector $\Pi_\s$, is given by
\be
    \rho_s=\frac{\Pi_\s\rho\Pi_\s}{\tr[\Pi_\s\rho]}\,.
\ee
By construction, the overlap of $\rho_\s$ with the aspired ideal, noise-free outcome $\rho_\mathrm{id}=\ket{\psi}\!\bra{\psi}$ increases as
\be
    \tr(\rho_\s\ket{\psi}\!\bra{\psi})=\frac{\tr[\rho\ket{\psi}\!\bra{\psi}]}{\tr[\Pi_\s\rho]}\geq\tr[\rho\ket{\psi}\!\bra{\psi}]\,.
\ee
The expectation value of an  observable $O$ with respect to the symmetry-projected state reads
\be
    \tr(O\rho_\s)=\frac{\tr[O\Pi_\s\rho\Pi_\s]}{\tr[\Pi_\s\rho]}=\frac{\tr[O_\s\rho]}{\tr[\Pi_\s\rho]}\,,
\end{equation}
where $O_\s=\Pi_\s O\Pi_\s$. 
If $[O,\Pi_\s]=0$, as is the case for gauge-invariant observables, this expectation value simplifies to 
\be
    \label{equ:exp_val_sym_commuting_obs}
    \tr(O\rho_\s)=\frac{\tr[O\Pi_\s\rho]}{\tr[\Pi_\s\rho]}\,.
\ee

For a lattice gauge theory with symmetry group $G$, the symmetry-preserving subspace S is given by invariance under the gauge transformations $\Theta_{g,v}\ket{\psi}=\ket{\psi}$ for all group elements $g\in G$ and all of the $n_\mathrm{v}$ vertices $v\in\mathrm{V}$. 
In the case of a finite (Abelian or non-Abelian) $G$, the projector $\Pi_\s$ takes the form
\be
    \label{equ:projector_gauge_theory}
    \Pi_\s=\prod_{v\in\mathrm{V}}\frac{1}{|G|}\sum_{g\in G}\Theta_{g,v}=\frac{1}{|G|^{n_\mathrm{v}}}\sum_{\mathbf{g}\in G^{\times\!n_\mathrm{v}}}\prod_{v\in\mathrm{V}}\Theta_{g_v,v}\,.
\ee
Inserting Eq.~\eqref{equ:projector_gauge_theory} into Eq.~\eqref{equ:exp_val_sym_commuting_obs} and employing the linearity of the trace yields
\be
    \label{equ:exp_val_sym_explicit}
    \tr(O\rho_\s)=\frac{\sum_{\mathbf{g}\in G^{\times\!n_\mathrm{v}}}\tr[\rho O\prod_{v\in\mathrm{V}}\Theta_{g_v,v}]}{\sum_{\mathbf{g}\in G^{\times\!n_\mathrm{v}}}\tr[\rho\prod_{v\in\mathrm{V}}\Theta_{g_v,v}]}\,.
\ee
With this, one may perform symmetry verification by post-processing measurement outcomes determined by the individual conservation laws and their correlations with the target observable.
That is, the expectation value of any gauge-invariant observable $O$ in the state projected into the symmetry-preserving subspace $\rho_\s$ can be obtained from a combination of measurements on the noisy state $\rho$. 

While this approach does not suffer from an exponentially decreasing number of valid shots, as does usual direct post-selection, the decompositions of Eq.~(\ref{equ:exp_val_sym_explicit}) employ on the order of $|G|^{n_\mathrm{v}}$ operators. In Sec.~\ref{sec:scaling}, we discuss the resource requirements in more detail.
Importantly, implementing this procedure does not require additional circuit elements in the form of ancillas or measurement gates and is not dependent on the commutation relations of the symmetry transformations.

\section{ Numerical benchmarks on the $D_3$ model}\label{sec:results}

To benchmark our gauge-protection protocols, we study the time evolution of a $2 + 1 \ D$ pure gauge lattice model with the dihedral $D_3$ group. This group represents the symmetries of an equilateral triangle and is the smallest non-Abelian group.
Having order $|D_3|=6$, it can be encoded exactly in existing qudit processors~\cite{ringbauer_NP2022} and thus presents an ideal setting to experimentally investigate the transition between strong- and weak-coupling limits. Moreover, dihedral models host non-Abelian topological order with universal anyons, making it also appealing in the framework of topological quantum computing~\cite{Shibo_ChinPhysLett2023,Andersen_Nature2023,Iqbal2024}.

$D_3$ has two one-dimensional irreducible representations ($j=e,p$) and one two-dimensional irreducible representation ($j=\tau$). The latter is the only faithful one, i.e., with a unique correspondence between group elements $g$ and matrices $D^\tau(g)$, up to a common unitary transformation.
Henceforth, we will work only with $j=\tau$.
A possible encoding of the group elements in a qudit with $d \geq 6$ is
\begin{equation}\label{eq:d3_mat}
    \ket{g} \Leftarrow D^\tau(g) =\begin{bmatrix}
    0 & 1 \\
    1 & 0
\end{bmatrix}^{\lfloor \frac{g}{3} \rfloor} 
\begin{bmatrix}
    e^{2\pi i/3} & 0 \\
    0 & e^{-2\pi i/3}
\end{bmatrix}^g \ ,
\end{equation}
where $g \in\{ 0, \ 1,\dots,5\} $ and $\lfloor\cdot\rfloor$ denotes the integer part. The two matrices in Eq.~\eqref{eq:d3_mat} belong to a faithful irreducible representation $D^\tau(g)$ of the generators  $s$ and $r$ of the group, see Appendix~~\ref{app:d3_details},  which represent respectively a reflection and a rotation by an angle $2\pi/3$.
The lattice geometry of our model consists of $2$ plaquettes and $2$ vertices with periodic boundary conditions (PBCs), so there are $4$ independent link variables $U_i$, see Appendix~\ref{app:d3_details} and Fig.~\ref{fig:C-Ops} for further details.

To illustrate the performance of the two non-Abelian gauge-protection schemes, we simulate the system evolution following a quantum quench. 
We initialize the system in a gauge-invariant state by setting every link in the trivial representation $j=e$, the ground state of $H_{\textrm{PG}}$, Eq.~\eqref{eq:HPG}, in the strong coupling limit ($g \gg 1 $). 
This state is given by a uniform superposition in the group element basis
\begin{equation}
    \ket{\psi_0} = \bigotimes_{\ell=1}^{L} \frac{1}{\sqrt{|D_3|}} \sum_{g\in D_3} \ket{g}_\ell \ ,
\end{equation}
with $L$ the number of links.
The system then evolves according to Eq.~\eqref{eq:gen_Ham}, whose Trotterized time-evolution is implemented according to Appendix~\ref{sec:Time_evo_imp}. 
To monitor the time evolution, we study the expectation value of the left plaquette operator $\mathcal{O}_{P1}$, which corresponds to
\begin{equation}
    \langle \mathcal{O}_{P1} \rangle = \langle\Re[\textrm{Tr}(U_0U_3U_0^{\dagger}U_2^{\dagger})]\rangle = -\frac{1}{2}a g^2 \langle H_B \rangle \,.
\end{equation} 
Since $ \langle H \rangle = \langle H_E \rangle + \langle H_B \rangle =\textrm{const} $, $ \langle \mathcal{O}_{P1} \rangle$ gives information on all terms constituting the Hamiltonian.

To mimic noise that affects any current-day quantum simulator, we include randomly drawn unitary operations $U_{\mathcal{E}(\gamma)}$ at each Trotter step. 
Here, $\gamma$ is a dimensionless parameter that determines the noise strength and the average distance of the random unitaries from the identity. 
Repeating the unitary evolution for $M$ independent trajectories allows us to estimate the ensemble-averaged dynamics subject to decoherence. 
We also checked for more specific, biased noise models, such as dephasing, and an increased system size, finding no qualitative difference regarding the efficacy of DPS and PSV (see Appendix~\ref{app:3p}).
More details on the group $D_3$ and the noise implementation are given in Appendices~\ref{app:d3_details},~\ref{app:error_model}. 

We numerically simulate the system dynamics with exact diagonalization in the full Hilbert space of dimension $6^L$. The code is available at~\cite{ballini_2025_15464248}.
In our simulations, we use a fixed Trotter step $\Delta t/a = 0.25$ and different noise strengths $\gamma$, and we set $\frac{1}{g^2} = 0.5$ in order to work in a scenario where $H_E$ and $H_B$ compete.

\subsection{Dynamical post-selection}
In this section, we evaluate the power of DPS to restore the desired gauge-invariant evolution of a noisy system. 
In the protocol we apply here, we protect only a single local symmetry after each noisy Trotter step, alternating vertices and group elements. For the given gauge group and system size, every local symmetry is checked on average every $t^{*} = 10\Delta t$. 
For Abelian theories, it has been shown that there exists a sharp threshold in the ratio between error rate and measurement frequency beyond which the gauge-protection regime (corresponding to a quantum Zeno effect) is left~\cite{wauters2024symmetryprotection}. 
Thus, one can expect the measurement frequency chosen here to be effective as long as the error strengths are not too large, whereas for higher $\gamma$ one should protect several symmetries after each Trotter step. 
In our simulations, the operations of the DPS protocol themselves are applied as if they do not introduce further errors, to highlight the physics of DPS.
Such additional errors could readily be modeled by an effectively larger noise strength $\gamma$. 

\begin{figure}
    \centering    \includegraphics[width=\linewidth]{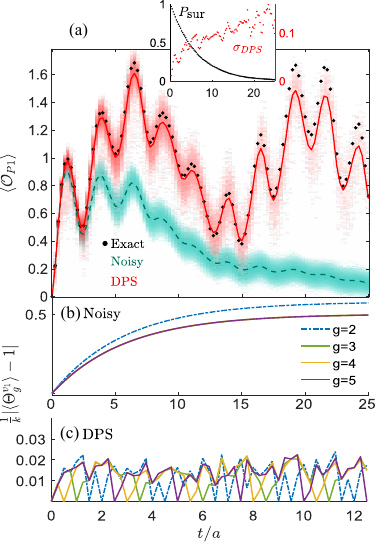}
    \caption{(a) Data clouds represent $\langle \mathcal{O}_{P1} \rangle$ at time $t/a$ on individual trajectories for DPS (red) and noisy (green) evolutions. The opacity reflects the fraction of trajectories retained after DPS. Solid lines represent ensemble averages. 
    The DPS ensemble average follows very well the error-free evolution (black dots).
    In the inset, we show the standard deviation of DPS data (red dots) and the fraction $P_{\textrm{sur}}$ of trajectories that always preserved the gauge symmetries (black dots). 
    (b-c) Gauge violations on vertex $v_1$ for the average of noisy evolutions without DPS (b) and for a single successful DPS trajectory (c). Reflection terms $g=3,4,5$ (solid lines) do not superimpose on a single noisy trajectory, while their averages do. Rotations are here represented by $g=2$ (dashed lines). Notably, the violation of a symmetry $\Theta_{g,v}$ can be partially suppressed also when other gauge charges are measured.  
    The system is evolved with $\gamma=0.2$, $1/g^2=0.5$, and $\Delta t/a=0.25$ for $M = 5000$ trajectories. The $x$-axis in panel (c) is zoomed in, compared to panels (a) and (b), to improve the readability. }
    \label{fig:Deph_PS}
\end{figure}
In Fig.~\ref{fig:Deph_PS}(a), we plot the results of $M = 5000$ trajectories affected by noise composed of random unitaries with a magnitude $\gamma = 0.2$, see Appendix~\ref{app:error_model}. 
DPS follows the exact dynamics closely up to the maximum considered evolution time ($t/a=25$). The protocol considerably improves over the noisy data, which strongly deviates from the error-free evolution already after $t/a = 3$. 
Due to the stochastic nature of the error protection protocol, the number of trajectories that preserve all tested gauge symmetries decreases exponentially in time, as shown by the black dots in the inset of Fig.~\ref{fig:Deph_PS}(a). 
Such a decrease is typical also for post-selection protocols based on data taken at the final time, which nevertheless have been used very successfully to restore symmetry-respecting time evolutions in Abelian theories~\cite{mildenberger2022probing, Nguyen_PRXQ2022, cochran2024stringZ2}.  
In the same inset, we plot the point-wise standard deviation of the successfully post-selected trajectories. Importantly, the standard deviation grows only slowly, reaching around $0.1$ (about $10\%$ of the mean) at the final considered time. 

To better visualize the effect of DPS, one can study the dynamics of the individual symmetry violations, which we define as 
\begin{equation} \label{eq:Gauge_Viol}
    \textrm{GV}_{g,v} = \frac{1}{k_g}|\langle \Theta_{g,v} \rangle - 1| \ .
\end{equation}
Here, $k_g$ is a normalization factor depending on $g$ such that $\textrm{GV} \in[0,1]$.
The Gauss' laws violation allows us to qualitatively compare post-selected and noisy dynamics and to assess the quality of the gauge-protection protocol. 
Figure~\ref{fig:Deph_PS}(c) displays the dynamics of $ \textrm{GV}_{g,v1}$ at the first vertex for a trajectory where the local symmetries are always successfully protected. 
Out of the six possible gauge transformations for each vertex, we can neglect the identity $g=0$ as it acts trivially. 
Moreover, we only monitor the violations corresponding to one rotation ($g=2$), since the $\mathbb{Z}_3$ subgroup of $\frac{2\pi}{3}$ rotations is cyclic, rendering conservation of the associated local symmetries equivalent. 
As shown in Fig.~\ref{fig:Deph_PS}(c), the DPS protocol repeatedly suppresses the gauge violations through the auxiliary measurements, and their growth between two successive measurements of the same symmetry operator remains limited. This pattern remains constant even for longer times, although the noisy evolution becomes increasingly decoupled from the exact one.
Moreover, gauge violations in Fig.~\ref{fig:Deph_PS}(c) decrease even when other symmetries are monitored, which indicates that gauge violations of different local charges are correlated.
The behaviour is markedly different in the bare noisy dynamics, shown in Fig.~\ref{fig:Deph_PS}(b).
Here, the ensemble-averaged violations increase until they reach a plateau, $0.5$ for reflections and $1/\sqrt{3}$ for rotations, corresponding to a fully mixed state\footnote{The precise asymptotic value depends on the definition in Eq.~\eqref{eq:Gauge_Viol}, but in general every charge sector is uniformly populated after a characteristic time depending on $\gamma$.}.

\begin{figure*}
    \centering
    \includegraphics[width=\linewidth]{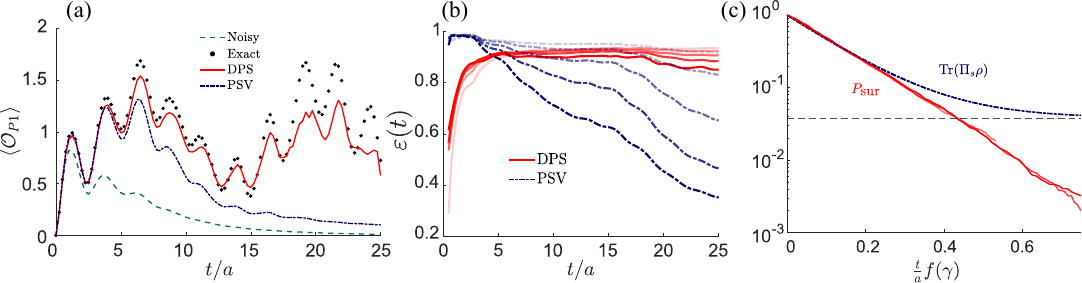}
    \caption{(a) Comparison between PSV (dot-dashed blue line) and DPS (solid red line) for relatively strong noise with $\gamma=0.3$, computed over 5000 trajectories. Both methods significantly improve over the average noisy evolution (dashed green line). PSV follows the exact Trotterized evolution (black dots) quantitatively well up to $t/a\simeq 6$, DPS even up to $t/a\simeq 17.5$. 
    (b) Protection efficacy, defined in Eq.~\eqref{eq:protection_e}, versus time, for increasing noise strength $\gamma = \{ 0.1,0.15,0.20,0.25,0.3\}$ (light to dark lines). PSV performs better initially but DPS remains resilient against noise for longer evolution times.
    (c) The survival probability of DPS trajectories and the gauge-symmetric projection of the density matrix in PSV, describing a signal-to-noise ratio, initially follow the same exponential decay. At late times, PSV reaches the value of the fully mixed density matrix, $\jdim \mathcal{H}_P /\jdim \mathcal{H}$ (horizontal dashed line).
    The rescaling $t/a\to \frac{t}{a} f(\gamma)$ collapses the data with different noise strengths. $f(\gamma)$ is defined in Eq.~\eqref{eq:p_err} and sets the effective error rate.
    Data for $1/g^2=0.5$ and $\Delta t/a= 0.25$.}
    \label{fig:dps_vs_psv}
\end{figure*}

\subsection{Comparing DPS and PSV}

Let us now explicitly compare the error-mitigation capacity of the two protocols described in Sec.~\ref{sec:protection}. First, we contrast the protection efficacy of both methods. After that, we discuss the relative scalings, which helps in assessing which method may be most suitable for a given platform.

\subsubsection{Protection efficacy}

As already anticipated in Fig.~\ref{fig:cover_pic}(b), both schemes successfully recover the exact dynamics up to times when the noise in the absence of gauge protection protocols has almost washed out any feature. 
Nevertheless, they are not identical, as PSV starts relaxing faster than DPS due to being oblivious to multiple gauge violations that combine into a gauge-invariant error.
This feature is increasingly evident if one tests them against stronger noise. 

In Fig.~\ref{fig:dps_vs_psv}(a), we compare the ensemble averages of the two protocols over $M=5000$ trajectories to the bare noisy evolution, for a high noise strength $\gamma=0.3$.
As above, in the DPS we perform one measurement per Trotter step. In the PSV scheme for the symmetry group $G=\mathrm{D}_3$ on $n=2$ vertices, Eq.~\eqref{equ:exp_val_sym_explicit} for the gauge-symmetric expectation value reduces to
\be   
    \label{equ:exp_val_sym_explicit_2}
    \tr(O\rho_\s)=\frac{\sum_{g_1,g_2\in \mathrm{D}_3}\tr[\rho O\Theta_{g_1,v_1}\Theta_{g_2,v_2}]}{\sum_{g_1,g_2\in \mathrm{D}_3}\tr[\rho \Theta_{g_1,v_1}\Theta_{g_2,v_2}]}\,.
\ee
Despite the strong noise, the two protocols produce very similar (and close to ideal) results up to $t/a\sim 6$, a time at which the noisy average has already lost any resemblance with the target dynamics. 
For $t/a \gtrsim 6$, PSV relaxes towards the infinite-temperature steady state while DPS continues to follow the target dynamics with small deviations until $t/a\sim 17.5 $. Eventually, DPS also deviates from the exact time evolution, all the while maintaining a rather satisfactory resemblance except for the very latest times we considered. 
Identifying and suppressing gauge violations already during the dynamics thus enhances the protection efficacy of DPS over PSV.
This better error mitigation capacity of DPS comes, however, at the cost of a larger number of entangling gates required for the measurement of the symmetry constraints through the ancillary qudits. In contrast, PSV just necessitates the addition of a layer of basis rotations at the end of the circuit in order to extract the relevant observables that constitute Eq.~\eqref{equ:exp_val_sym_explicit_2}.

To get a better quantitative understanding of how the performances of the two protocols depend on the noise strength $\gamma$, we consider the cumulative error over time
\begin{equation}\label{eq:errcum}
    \mathcal{E}^{*}(t) = \sum_{t_n \leq t}|\langle H_B(t_n) \rangle_{*} - \langle H_B (t_n)\rangle_{\rm{\rm id}}| \ ,
\end{equation}
where $t_n$ labels the discrete time slices, $\langle H_B \rangle_{{\rm id}}$ is the magnetic energy measured after an ideal error-free Trotterized evolution, and the sub-/superscript $*\in\{\textnormal{DPS, PSV}\}$ represents the use of one of the two methods on top of a noisy evolution. 
We define the protection efficacy as the normalized difference between cumulative errors with and without gauge protection
\begin{eqnarray}\label{eq:protection_e}
    \varepsilon(t) = 1-\mathcal{E}^{*}(t)/\mathcal{E}^{\textrm{Noisy}}(t) \ .
\end{eqnarray}
When $\varepsilon = 0$, the gauge protection protocol does not improve over the bare noisy evolution, while $\varepsilon=1$ corresponds to perfect error mitigation matching the error-free dynamics. 

We compare the protection efficacy of DPS and PSV in Fig.~\ref{fig:dps_vs_psv}(b) for different values of $\gamma \in [0.1,0.3]$.
PSV displays an advantage at short times as it incorporates all gauge transformations {\em at once}. In contrast, our implementation of DPS only monitors one local symmetry per Trotter step, thus requiring a longer time to properly suppress the gauge violations.
On the other hand, DPS is more stable from intermediate to long times, consistent with Fig.~\ref{fig:dps_vs_psv}(a), and is less affected by the increasing noise strength.
Our results also suggest that increasing the error rate in DPS due to imperfect mid-circuit measurements will not impact qualitatively its protection efficacy, as the algorithm performance is stable from weak to intermediate-strength noise.

\subsubsection{Resource requirements}
\label{sec:scaling}

A natural question is whether any of the two methods offers a practical advantage over the other, in situations where they present similar error-mitigation quality. Roughly speaking, PSV is more flexible and can be used in any architecture, whereas DPS seems beneficial in platforms where adequate mid-circuit measurement operations are available.  To be able to make more precise statements, it is useful to consider the scaling of both methods.

To implement DPS with full symmetry verification after each Trotter step for $n_\mathrm{v}$ vertices, one needs at most $|G| n_t n_\mathrm{v}$ single shot mid-circuit measurements during a single run. 
Since these measurements concern local operators, those acting on non-neighboring vertices can be parallelized. Hence, the resulting increase in the number of circuit layers does not depend on the system size.
The above scaling with $|G|$ can be further improved by measuring only an optimal generating set of $G$ (e.g., for $D_3$, the reflection and the rotation with smallest angle are already enough to ensure full protection). 
Further, to circumvent costly mid-circuit readout and reset~\cite{Baumer_PRXQ2024, hashim2024quasiprobabilistic, Ivashkov_PRXQ2024}, one may also employ an array of ancillas whose measurement is delayed to the end of the circuit.
The additional spatial (ancillary qudits) and logical resources (two-qudit gates coupling the ancillas to the system) need to be sufficiently precise, as otherwise they would wash out the benefits of gauge protection.
In contrast, PSV needs only basis rotations at the end of the simulations to measure the desired operators, making a single run less demanding.

As for any similar mitigation method, the present algorithms have a sampling overhead that is exponential in the simulated time, i.e., number of Trotter steps,  and system size, which, however, becomes manifest differently in the two protocols:
in DPS, the dynamics will eventually be dominated by the proliferation of gauge-violating trajectories, meaning an ever smaller number of runs can be successfully retained as the size and depth of the circuit increases; in comparison, the decreasing weight of the group-symmetric density matrix $\Pi_s \rho$, leading to a small denominator in Eq.~\eqref{equ:exp_val_sym_explicit}, limits PSV performance.

Focusing on the scaling in the number of Trotter steps at a fixed system size, the survival rate of DPS, i.e., how many trajectories never incur a measurement yielding a ``wrong'' result, can be estimated assuming that the amplitude $\beta$ in Eq.~\eqref{eq:trott_step_DPS} is  $\propto \gamma$. In that case, the error channel populates the gauge-violating sector with a rate $p_{\textrm{err}} \sim \gamma^2$. 
More precisely, we found that the error rate is associated with the average Hilbert--Schmidt inner product $\overline{\langle U_{\mathcal{E}(\gamma)},\mathbb{1}\rangle}_\mathrm{HS}$ between the identity (i.e., zero noise) and the random unitaries $U_{\mathcal{E}(\gamma)}$ drawn according to the chosen noise model,
\begin{equation}\label{eq:p_err}
    p_{\textrm{err}}\simeq f(\gamma)= 1-\frac{\overline{\langle U_{\mathcal{E}(\gamma)},\mathbb{1}\rangle}_\mathrm{HS}}{{\rm dim}\mathcal{H}} = 1 - \nep^{-\frac{\gamma^2}{2}}\left(1-\frac{2}{{\rm dim}\mathcal{H}}\right).
\end{equation}
At the lowest order in $\gamma$, indeed $p_{\textrm{err}} \sim \gamma^2$, while the expression for $\overline{\langle U_{\mathcal{E}(\gamma)},\mathbb{1}\rangle}_\mathrm{HS}$ is given by the error model described in Appendix~\ref{app:error_model}.
The probability of never incurring any gauge violations after $n_t=t/\Delta t$ Trotter steps is therefore
\begin{eqnarray}
    P_s(t)=(1-p_{\rm err})^{n_t} \sim \nep^{-\frac{t}{\Delta t}f(\gamma)} \ .
    \label{eq:Ps}
\end{eqnarray}
In Fig.~\ref{fig:dps_vs_psv}(c), we show the survival probability $P_s(t)$ for DPS as well as the weight of the symmetrized density matrix $\tr[\Pi_s \rho]$ using PSV for $\gamma\in[0.1,0.3]$.
The rescaling $t/a \to \frac{t}{a} f(\gamma)$ suggested by Eq.~\eqref{eq:Ps} leads to a perfect collapse of the data and highlights that both methods require an exponentially scaling overhead to properly estimate the gauge-invariant expectation values.
The main difference between the two protocols is a saturation occurring for PSV at long times, since $\tr[\Pi_\s \rho] \to {\rm dim}{\mathcal{H}_P}/{\rm dim}(\mathcal{H})$ as a consequence of the steady state being fully mixed.
The time around which $\tr[\Pi_\s \rho]$ deviates from the exponential decay ($\frac{t}{a}f(\gamma)\approx 0.3$ in Fig.~\ref{fig:dps_vs_psv}c) coincides with the time around which the reconstructed PSV signal deviates significantly from the ideal data (for $\gamma=0.3$, $t/a\approx 7$, see Fig.~\ref{fig:dps_vs_psv}a).

For PSV, naively one could expect that one would need to construct all expectation values in Eq.~\eqref{equ:exp_val_sym_explicit} explicitly, which would imply $2|G|^{n_\mathrm{v}}$ independent measurement settings.
The actual cost can be considerably lower.
First, a gauge transformation $\prod_{v\in\mathrm{V}}\Theta_{g_v,v}$ and its correlator with a gauge-invariant observable $O$ may be measured simultaneously.
Further, one can employ the group commutation structure to identify groupings of commensurable or related contributions of different group elements in Eq.~\eqref{equ:exp_val_sym_explicit}, reducing the number of different measurements needed.
Generally, the question of how many independent measurements are required may be formalized as finding a minimum clique cover on a graph representing the commutation structure of the operators, which here is given by the commutators within $G^{\times\!n_\mathrm{v}}$.
In the numerical example of the above section for $n_\mathrm{v}=2$ vertices and the gauge group $G=D_3$, this allows one to reduce from the $2\cdot6^2=72$ individual expectation values appearing in Eq.~(\ref{equ:exp_val_sym_explicit_2}) to $4^2=16$ cliques of simultaneously measured operators. 
A further efficiency improvement could be achieved by restricting the full projector to one that only acts in the vicinity of the local observables of interest.

Additionally, one can estimate Eq.~\eqref{equ:exp_val_sym_explicit} through a random sampling of the group element combinations $\lbrace g_1,\dots, g_{n_\mathrm{v}} \rbrace $. For every sample, one simultaneously measures the prescribed gauge transformation operator and its correlator with the gauge-invariant target observable (or cardinality-weighted cliques thereof). Analogous considerations for the qubit case \cite{cai2021quantum, mcclean2020decoding, huggins2021efficient} reveal such estimators to require a sampling overhead scaling as $\tr[\Pi_\s\rho]^{-2}$. This overhead can generally be determined by analyzing the increase in the variance of the new estimator compared to the unmitigated, noisy one. The variance of a ratio of expectation values, such as those estimators, is then dominated quadratically by the inverse of the denominator.
The decrease of $\tr[\Pi_\s\rho]$ with circuit depth and system size is governed by the strength of (gauge-invariance breaking) error rates, and therefore the required sampling cost immediately benefits from any quantum hardware developments. 
In particular, for a given shot budget and error rate, one has a window of simulated time within which controlled sampling of the gauge-invariant expectation value is possible.

As a final remark, the feasibility and quality of the mid-circuit measurements are a fundamental requirement to implement DPS. 
While qudit hardware is still in a rapidly developing stage and the various platforms have yet to demonstrate their implementation, mid-circuit measurements are routinely used in qubit processors for a number of purposes, e.g., state preparation \cite{Iqbal2024} or long-range entangling operations \cite{Baumer_PRXQ2024}.
The extension to qudit devices requires some care, as the readout process becomes slower and more delicate because of the many levels involved, but faces no fundamental obstacle. 
In addition, efficient mid-circuit monitoring of ancilla qubits is a crucial building block for any error-correcting code and, as such, is actively pursued in many platforms aiming towards fault-tolerance.

\section{Conclusions}\label{sec:conclusion}

In this work, we have discussed symmetry-verification frameworks modified to work with quantum simulations of non-Abelian lattice gauge theories in noisy devices, where fault-tolerant computation is beyond reach and error mitigation strategies are essential to extract meaningful physical information. 
The non-commutative nature of the gauge transformations hinders their simultaneous projective measurement and prevents the use of direct ancilla-free post-selection of experimental runs, requiring a novel approach to protect non-Abelian gauge invariance.
Focusing on qudit-based architectures, we have discussed two protocols: the first, termed dynamical post-selection (DPS), preserves gauge invariance through frequent mid-circuit measurements of ancillary qudits encoding the local gauge charges. The second, post-processed symmetry verification (PSV), involves estimating group-symmetrized expectation values by measuring correlations between the target observable and Gauss' law operators.

We have applied these methods to the dihedral group $D_3$, the smallest discrete non-Abelian group, which allows for an exact encoding on current qudit hardware.
Both approaches significantly improve the Trotterized time evolution in the presence of noise. 
As they offer distinct advantages and face different constraints, these two tools neatly complement each other.

DPS is remarkably resilient against noise and the measurement strategy can be easily adapted to suit specific needs. For instance, the density of mid-circuit measurements can be tuned according to the system studied and the error channel expected on a specific hardware.
The only requirement is that the measurement frequency must be above the quantum Zeno threshold for error mitigation in the system, otherwise, the system would enter an unprotected regime with random dynamics jumping between different gauge sectors~\cite{wauters2024symmetryprotection}. 
The two-qudit gates used in the DPS circuit also contribute with an additional error rate. Nevertheless, as long as the number of additional two-qudit gates is sufficiently smaller than the ones already required for the Hamiltonian time evolution, this does not constitute a decisive limiting factor.
 On a qudit-based architecture, one needs $6(|G|-1)$ control-permutations per Trotter step per plaquette, using the strategy sketched in Appendix \ref{sec:Time_evo_imp}. 
In contrast, checking one local symmetry $\Theta_{g,v}$ requires $4(|G|-n_g)$ control-permutations, where $n_g$ is the number of states that transform trivially under the group element $g$.
For the two-plaquette system considered above, this leads to 60 control-permutations for each Trotter step and 8 (10) for a reflection (rotation) symmetry check\footnote{These number reflect the fact that each vertex has only three links in our geometry and the trivial subspace for the vertical links has dimension 4 for both reflection and rotations, see Appendix \ref{app:d3_details}.}.
This advantage shows a promising window of opportunity for future experimental studies on qudit-based hardware during the NISQ era.
Moreover, the resource requirements can be mitigated by choosing an optimal subset of group elements for the symmetry measurements: gauge violations on a single vertex $v$ can be fully suppressed by monitoring only a suitably chosen generating set of $G$.

Conversely, the circuit depth for PSV increases at most by a basis rotation above the ``bare'' time evolution in order to measure the correlators between the desired observables and symmetry transformation operators. 
Instead of explicitly measuring all of these individually, a significant reduction of required measurements can be achieved by groupings of commensurable operators and sampling strategies. Moreover, one can restrict the projector to enforce only a local subset of the symmetries.
Further, PSV retains the entire ensemble of shots taken, unlike direct post-selection procedures, and therefore does not alter the error channel acting on the circuits; it may thus be readily combined with other error mitigation methods explicitly depending on information about the present error channel.

Our findings contribute to expanding the toolkit for quantum simulations on noisy quantum hardware, focusing on techniques to protect non-Abelian gauge symmetries. We hope these results inspire further research into error-mitigation protocols for such NISQ devices.

\section*{Acknowledgements}
We thank K. Mato, A. Biella, P. Silvi, E. Tirrito, M. Ringbauer, G. Clemente, R. Costa de Almeida, and M. Burrello for fruitful discussions.
This project has received funding from the European Union’s Horizon Europe research and innovation programme under grant agreement No 101080086 NeQST, from the Italian Ministry of University and Research (MUR) through the FARE grant for the project DAVNE (Grant R20PEX7Y3A).
P.H. has further received funding from Fondazione Cassa di Risparmio di Trento e Rovereto (CARITRO) through the project SQuaSH - CUP E63C24002750007, from the Swiss State Secretariat for Education, Research and Innovation (SERI) under contract number UeMO19-5.1, from the QuantERA II Programme through
the European Union’s Horizon 2020 research and innovation programme under Grant Agreement No 101017733, from the European Union under NextGenerationEU, PRIN 2022 Prot. n. 2022ATM8FY (CUP: E53D23002240006), from the European Union under NextGenerationEU via the ICSC – Centro Nazionale di Ricerca in HPC, Big Data and Quantum Computing.
Views and opinions expressed are however those of the author(s) only and do not necessarily reflect those of the European Union or the European Commission. Neither the European Union nor the granting authority can be held responsible for them.
This work was supported by Q@TN, the joint lab between the University of Trento, FBK—Fondazione Bruno Kessler, INFN—National Institute for Nuclear Physics, and CNR—National Research Council.

\section*{Data availability statement}
All numerical data and the code we used in this work are available at~\cite{ballini_2025_15464248}.

\appendix

\section{Trotterized time evolution of a discrete LGT}\label{sec:Time_evo_imp}
Here, we briefly review and adapt to our implementation the main ingredients for digital quantum simulations of non-Abelian LGT~\cite{Lamm_PRD2019}.
To perform a single Trotter step 
$
    e^{-iH_E \Delta t} e^{-i H_B \Delta t} \ ,
$
we need a set of {\em primitive} gates: high-level constructs that allow us to perform the time evolution corresponding to the electric and magnetic terms.

On qudit hardware where we can encode the gauge field exactly, the implementation of the unitary operator associated with $H_E$ is trivial. One can diagonalize the single-qudit Hamiltonian derived by Eq.~\eqref{eq:Tk} and feed the eigenvalues $\{\epsilon_j\}$ to a phase gate 
\begin{equation}\label{eq:phase_el}
    \mathfrak{U}_{\rm ph}(\epsilon_j \Delta t) ={\rm diag}\left(\nep^{-i\epsilon_j \Delta t} \right) \ ,
\end{equation}
with $j$ labeling the irreducible representations; each eigenvalue has at least degeneracy $\textrm{dim}(j)^2$.
In the group element basis, $\mathfrak{U}_{\rm ph}(\epsilon_j \Delta t)$ has to be rotated according to the generalized Fourier transform $\mathfrak{U}_{\textrm{QFT}} = \langle g|jmn\rangle $ defined in Eq.~\eqref{eq:GFT}. So, the evolution of the electric Hamiltonian $H_E$ reads
\begin{eqnarray}
\nep^{-iH_E\Delta t} = \bigotimes_l \mathfrak{U}_{\textrm{QFT}, l} \mathfrak{U}_{{\rm ph}, l}(\mathbf{\epsilon}_j \Delta t) \mathfrak{U}^{-1}_{\textrm{QFT}, l} \ .
\end{eqnarray}

\begin{figure}
    \centering
    \includegraphics[width=\columnwidth]{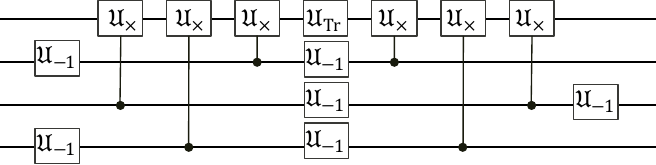}
    \caption{Circuit decomposition implementing the time evolution generated by a plaquette term in $H_B$. It has to be applied to every plaquette of the theory. Each line represents the quantum register of a group element subspace. This architecture is analog to the one presented in Ref.~\cite{Lamm_PRD2019}.}
    \label{fig:mult_circ}
\end{figure}

The unitary operator generated by the four-body interaction described by the magnetic term $H_B$ in Eq.~\eqref{eq:h_v} is more complicated to implement. A possible circuit decomposition~\cite{Lamm_PRD2019}, depicted in Fig.~\ref{fig:mult_circ}, is a generalization of a standard four-body Pauli-$Z$ interaction for qubits: three qudit control-multiplication gates acting on the fourth, onto which subsequently a {\em trace} gate is applied. Finally, the state of the four qudit registers is restored by applying the inverse of the multiplication gates. 
The procedure also requires the application of the inversion gate, which maps an element of the group $g$ into its inverse $\mathfrak{U}_{-1}\ket{g}=\ket{g^{-1}}$. 
The trace gate is diagonal in the group-element basis and implements the parametric operation
    \begin{eqnarray}
        \bra{g'} \mathfrak{U}_{\rm Tr} (\Delta t) \ket{g} = \nep^{i\Delta t {\rm Tr} D^j(g)} \delta_{g'g} \ .
    \end{eqnarray}
The control-multiplication gate is defined as
\begin{eqnarray}
        \mathfrak{U}_X(\ket{g} \otimes \ket{h}) = \ket{g}\otimes \Theta^L_g\ket{h} = \ket{g} \otimes \ket{gh} \ , 
\end{eqnarray}
where the first qudit is the control and the second is the target. This is the most costly operation in quantum hardware, as it applies $|G|$ different operations (one of which, the identity, is trivial) depending on the state of the control qudit. 
The control-multiplication gate $\mathfrak{U}_X$ is the only two-qudit unitary required for the time evolution. 
To distinguish between $\mathfrak{U}_X$ and a group-multiplication controlled by specific group elements, we denote the latter as control-permutation. A single control-multiplication is thus a sequence of $|G|-1$ control-permutations. 

On qubit hardware, in contrast, the QFT becomes a set of entangling gates. Its efficient implementation, in general, is not known, but a lot of effort has been made to find the optimal QFT for certain physically relevant non-Abelian groups~\cite{MurairiPRD2024}. One can gain some insight from $D_4$, whose QFT can be efficiently implemented using two entangling gates~\cite{Alam_PRD2022}. We can estimate that using qubits instead of qudits for $D_4$ would require $2\times L \times 2 $ additional entangling gates for each Trotter step, where $L$ is the number of links in the lattice, the first factor $2$ is the number of entangling gates for every QFT, and the second factor $2$ represents the two QFTs we have to perform on every link at every Trotter step. In Ref.~\cite{GonzalezCuadra_PRL2022}, the advantage of implementing the multiplication gate on qudits instead of qubits is analyzed for $\mathbb{Q}_8$, the quaternion group that is a non-Abelian subgroup of $SU(2)$ with cardinality $8$. In particular, they estimate a reachable fidelity of $99.6\%$ for qudits versus $21.4\%$ for qubits.

\section{Inclusion of matter fields in the LGT}\label{sec:matter}
Although non-Abelian pure gauge theories already display many interesting phenomena (prominent examples being confinement and glue balls in QCD \cite{gattringer2010quantum}), nature is not only pure gauge. 
In the interest of being self-contained, this appendix shows how matter fields can be straightforwardly included in the present gauge theory. 

$D_3$ has only one faithful irreducible representation of dimension 2, corresponding to the number of colors, which we label with $a,b$. The basis states for a single matter site are thus $\{\ket{0}, \ket{a}, \ket{b}, \ket{ab}\}$, denoting absence of particles, presence of one of either color $a$ or $b$, or presence of both colors.
Within a Kogut--Susskind description of a  $2+1$ dimensional theory~\cite{kogut1975, Zohar_PhilTransA2021}, fermionic fields on the vertices, labeled by $\textbf{x}=(x_1,x_2)$, are governed by the following two additional terms in the Hamiltonian 
\begin{equation}
\begin{split}
    & H_{\textrm{m}} =  m \sum_{\textbf{x}}\sum_{\zeta\in\{a,b\}} (-1)^{x_1+x_2} \hat{n}_{\textbf{x},\zeta}, \\
    & H_J = J \sum_{\textbf{x};i\in\{1,2\}} \ \sum_{\zeta,\zeta'\in\{a,b\}}(\psi^{\dagger}_{\textbf{x},\zeta} U^{\zeta,\zeta'}_{\textbf{x},\textbf{x}+\hat{e}_i} \psi_{\textbf{x}+\hat{e}_i,\zeta'} + \textrm{h.c.}), 
\end{split}
\end{equation}
where $\textrm{m}$ is the particle mass, $(-1)^{x_1+x_2}$ takes into account the staggering of the mass, addressing the fermion doubling problem, $\hat{n}_{\textbf{x},\zeta}$ is the occupation number operator for lattice site $\textbf{x}$ and color $\zeta$, and $(\psi^{\dagger}_{\textbf{x},\zeta} U^{\zeta,\zeta'}_{\textbf{x},\textbf{x}+\hat{e}_i} \psi_{\textbf{x}+\hat{e}_i,\zeta'} + \textrm{h.c.})$
describes the gauge-invariant hopping term between neighboring vertices, 
where we labeled the links by the indices of the vertices they connect. Further, $U^{\zeta,\zeta'}_{\textbf{x},\textbf{x}+\hat{e}_i}\ket{g}_{\textbf{x},\textbf{x}+\hat{e}_i}=D^\tau_{\zeta,\zeta'}(g)\ket{g}_{\textbf{x},\textbf{x}+\hat{e}_i}$ is the group connection in the faithful representation $\tau$ and $\psi_{\textbf{x},\zeta}^{\dagger}$ and $\psi_{\textbf{x},\zeta}$ are the creation respectively annihilation operators for a fermion of color $\zeta$ on site $\textbf{x}$.

When matter is included, the local gauge transformation has to act also on the matter field at the vertex $n$, meaning Eq.~\eqref{eq:gauss_general} is modified to 
\begin{equation}
    \Theta_{g,v} = \prod_o \Theta^L_{g,o} \prod_i \Theta^R_{g,i} \Theta_g^Q\,.
    \label{eq:fulltra}
\end{equation}
Here, $\Theta_g^Q$ is the charge transformation, defined as
\begin{equation}
    \Theta_g^{Q} = e^{i\psi^{\dagger}_\zeta q_{\zeta\zeta'}(g)\psi_{\zeta'}}\textrm{det}(g^{-1})^N \ ,
    \label{eq:chargetrafo}
\end{equation}
where 
\begin{equation}
    q(g) = -i\textrm{log}(D^\tau(g))\ ,
\end{equation}
and $N=0$ for a vertex in the even sublattice and $N=1$ for the odd one. 
In appendix~\ref{app:d3_details}, we show how the DPS and PSV protocols can be extended to include matter fields.

\section{$D_3$ lattice gauge theory}\label{app:d3_details}

$D_3$ is the dihedral group of order $|D_3| = 6$ and represents the symmetries of an equilateral triangle.
It is the smallest non-Abelian group, providing an ideal and relatively simple benchmark for implementing quantum simulations and gauge protection schemes. 
Each group element can be written as $g=s^k r^h$, where $k\in\{0,1\}$ and $h\in\{0,1,2\}$, while $s$ and $r$ are a reflection and the $2\pi/3$ rotation respectively. 
A faithful unitary representation for $s$ and $r$ reads
\begin{eqnarray}
    D^\tau(s) = \begin{bmatrix}
    0 & 1 \\
    1 & 0
\end{bmatrix} , \  D^\tau(r) =\begin{bmatrix}
    e^{2\pi i/3} & 0 \\
    0 & e^{-2\pi i/3}
\end{bmatrix}.
\end{eqnarray}
We can encode the group elements in a qudit with $d \geq 6$ as $\ket{g}=\ket{s^{\lfloor \frac{g}{3} \rfloor} r^g}$,
where $g \in\{ 0, \ 1,\dots,5\} $ and $\lfloor\cdot\rfloor$ denotes the integer part, such that $\ket{1}, \ket{2}$ are associated to rotations and $\ket{3},\ket{4},\ket{5}$ to reflections. $\ket{0}$ is associated with the identity operator.

\begin{figure}
    \centering
\includegraphics[width=\linewidth]{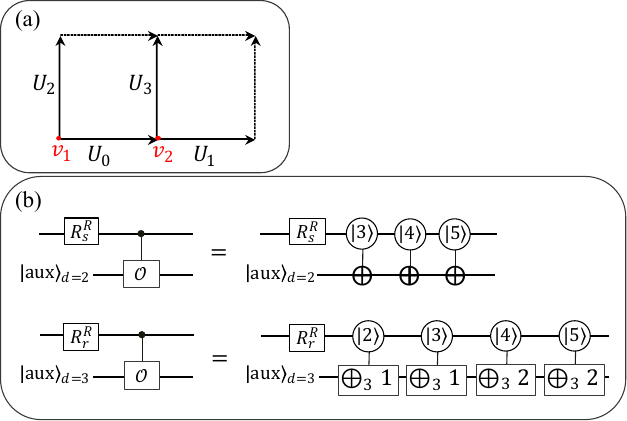}
    \caption{A possible implementation of the $C$-$\mathcal{O}$ after $R^R$, both for reflections (labeled with $s$, top circuit) and for rotations (labeled with $r$, bottom circuit). $d=2,3$ defines the dimension of the auxiliary qudit and $\bigoplus_3$ indicates the sum modulo $3$ ($\mathbb{Z}_3$ cyclic permutations). We add $1$ or $2$ depending on the qudit state controlling the operation.}
    \label{fig:C-Ops}
\end{figure}

The lattice geometry of our toy model consists of $2$ plaquettes and $2$ vertices with periodic boundary conditions (PBCs), see Fig.~\ref{fig:C-Ops}(a), so there are $4$ independent link variables.
The full Hilbert space has dimension $\textrm{dim}(\mathcal{H}) = 6^L = 1296$, for $L=4$ links but
the physical gauge sector space is considerably smaller, with dimension~\cite{Mariani_PRD2023} $\textrm{dim}(\mathcal{H_P}) = \sum_C \left[\frac{|G|}{|C|}\right]^{L-V} = 49$. $C$ represents conjugacy classes of the group and $V$ is the number of vertices.
The gauge transformations for the two vertices $v_1$ and $v_2$ are
\begin{align}\label{eq:Theta_gv1}
    \Theta_{g,v1} = \Theta_g^L \otimes \Theta_g^R \otimes\Theta_g^L\Theta_g^R\otimes\mathbb{1}_6 \ , \\
    \Theta_{g,v2} = \Theta_g^R \otimes \Theta_g^L \otimes \mathbb{1}_6 \otimes\Theta_g^L\Theta_g^R \ , \label{eq:Theta_gv2}
\end{align}
where we dropped the link indices as they are identified by the order in the tensor product.
The terms  $\Theta_g^L \Theta_g^R$ reflect the ``double'' directionality of the vertical links, which are both ingoing and outgoing from the vertices due to PBCs. 
However, this modifies neither the eigenvalues of $\Theta_{g,v}$ nor the possibility of diagonalizing them with single-qudit rotations.

In the DPS protocol, the nature of mixed-dimensional controlled operators $C$-$\mathcal{O}$ introduced in Fig.~\ref{fig:vertex_circuit}(b), as well as the dimension of the auxiliary qudit, depends on the group element we decide to protect. 
For the reflections $g\in\{3,4,5\}$, the $\Theta_s^{R(L)}$ operators have $2$ distinct eigenvalues,  
\begin{equation} \label{eq:ref_eigs}
    \sigma^{R(L)}_{g\in\{3,4,5\}} = \{1,1,1,-1,-1,-1\} \ ,
\end{equation}
so a qubit is enough to encode both of them. 
On the other hand, for non-trivial rotations, $g\in\{1,2\}$, the operators $\Theta_r^{R(L)}$ have $3$ distinct eigenvalues,  
\begin{equation}\label{eq:rot_eigs}
\begin{split}
    \sigma^{R(L)}_{g\in\{1,2\}} = \{1,1,\exp{(2\pi i/3)},\exp{(2\pi i/3)}, \\
    \exp{(-2\pi i/3)},\exp{(-2\pi i/3)}\} \ ,
\end{split}
\end{equation}
meaning we need a qutrit auxiliary register.

Figure~\ref{fig:C-Ops}(b) shows a possible way to implement the $C$-$\mathcal{O}$ operators after the change of basis on the computational qudits.
When the protected symmetry is a reflection, we encode the information of its eigenvalues in the auxiliary qudit. This is done by applying a $X$ (NOT) operator when the eigenstate encoded on the controlling qudit has eigenvalue $-1$. 
When the protected symmetry is a rotation, its eigenvalues are encoded through a sum modulo $3$ on the auxiliary qutrit: 
we add $1$ ($2$) to the ancilla when the eigenstate encoded on the controlling qudit has eigenvalue $\exp(2\pi i/3)$ ($\exp(-2\pi i/3)$). 
The choice, however, is not unique and the order of operators can be arranged as is most convenient. 
 When the operator acting on the link variable is $\Theta_g^L \Theta_g^R$, $C$-$\mathcal{O}$ operators need two controlled gates since the eigenvalues are $\sigma_{g\in\{3,4,5\}}^{RL} = \{1, 1, 1, 1,-1, -1\}$ and $\sigma_{g\in\{1,2\}}^{RL} = \{1, 1, 1, 1,\exp{(2\pi i/3)}, \exp{(-2\pi i/3)}\}$ for reflections and rotations respectively.

\subsection*{DPS and PSV with qubits}
Even though in this work we explore the application of DPS and PSV on a qudit-based hardware, these methods can be used on a qubit-based platform as well. In DPS, employing qubit-based hardware would increase the number of entangling gates to perform the change of basis in Fig.~\ref{fig:vertex_circuit}, an operation that on qudits can be implemented with single-qudit rotations. The resources needed depend on the group structure and on the native gates of the qubit hardware. Furthermore, the number of entangling gates for the encoding operation in Fig.~\ref{fig:C-Ops} increases as well. For example, for $D_3$, the verification of the rotational symmetries will require entangling two auxiliary qubits with the $12$ qubits used to encode the degrees of freedom around a vertex. For the PSV, using qubits instead of qudits requires an increasing number of entangling gates for the change of basis before the measurement.
 
\subsection*{DPS and PSV including matter}
To apply the DPS protocol in the presence of matter fields, one has to also take into account the eigenvalues of the charge transformation $\Theta_g^{Q}$ as given in Eq.~\eqref{eq:chargetrafo}.
As discussed in Appendix~\ref{sec:matter}, the matter fields can be encoded in a qudit with local dimension $4$. 
To perform the mid-circuit measurement, one should rotate the matter fields into the basis that diagonalizes $\Theta_g^Q$ and then perform a controlled operation acting on an ancilla analogously to the ones depicted in Fig.~\ref{fig:C-Ops}(b) for the gauge fields.
In our $D_3$ model, the eigenvalues are the same as the ones in Eq.~\eqref{eq:ref_eigs} and~\eqref{eq:rot_eigs}, with different multiplicity because the dimension of the matter Hilbert space is smaller than the gauge one,
\begin{equation}
\begin{split}
    &\sigma^Q_{g\in\{3,4,5\}} = \{1, 1, -1, -1 \}\\
    &\sigma^Q_{g\in\{1,2\}} = \{1, 1, \exp{(2\pi i/3)}, \exp{(-2\pi i/3)}\}.
\end{split}
\end{equation}
Thus, the controlled operators can encode the eigenvalues of $\Theta_g^Q$ on the same auxiliary qudits as shown in Fig.~\ref{fig:C-Ops}.
To extend PSV to theories including matter, one has to include charge transformations, Eq.~\eqref{eq:chargetrafo}, in the local gauge transformations. This does not increase the number of different operators in the decompositions of Eq.~(\ref{equ:exp_val_sym_explicit}), remaining $|G|^{n_\mathrm{v}}$, where $n_\mathrm{v}$ is the number of vertices.

\begin{figure}[t!]
    \centering
    \includegraphics[width=0.85\linewidth]{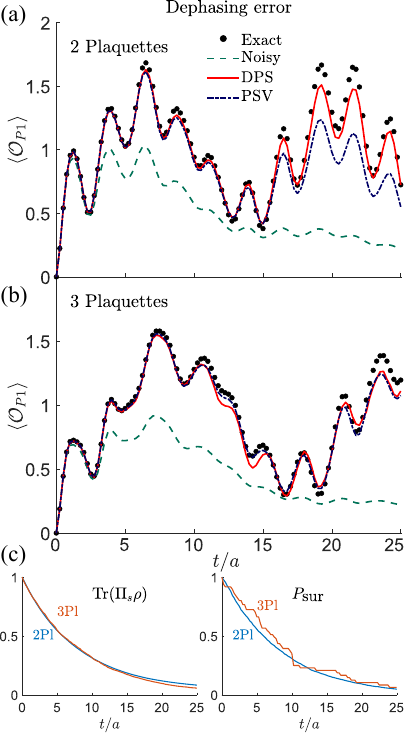}
    \caption{Comparison between the exact, noisy, DPS, and PSV time evolution in presence of dephasing noise for (a) two plaquettes, averaged over 5000 independent trajectories, and (b) three plaquettes, averaged over 145 independent trajectories. 
    (c) At constant local error rate, the weight of the gauge-invariant part of the state for PSV and the survival probability in the case of DPS remain comparable (for DPS, there is still a single measurement per Trotter step). 
    Thus, from 2 to 3 plaquettes the protection quality of DPS only slightly decreases, due to the reduced measurement density, while maintaining a constant resource cost. For PSV, by contrast, the protection quality even increases, at the cost of 6 times more observables to be measured, and thanks to the smaller ratio between the dimensions of the physical sector and the total Hilbert space.
    For the two-plaquette system, $\gamma=0.2$ as previously, while for the three-plaquette system we rescaled to the corresponding $\gamma=0.245$, as explained in the text.}
    \label{fig:3_plaquettes}
\end{figure}

\section{Error model}\label{app:error_model}
To introduce errors in the simulations and test the gauge protection capability of our protocols,
we consider a noise channel generated by random unitaries close to the identity. While this choice maintains unitarity in the time evolution on a single run (trajectory), the ensemble average results in an incoherent error channel.

Several techniques can numerically generate random unitaries close to the identity. Here, we implement a method based on Householder transformations~\cite{Horn_Johnson_2012}.
These are unitary transformations describing the reflection about a hyperplane that contains the origin. The matrix operator describing the Householder reflection defined by its normal vector $\boldsymbol{v}$ reads $\mathcal{R}=\mathbb{1}-2\boldsymbol{v}\boldsymbol{v}^{\dagger}$, where $\boldsymbol{v}$ in our case is a random complex vector whose elements are extracted from a normal distribution with unit variance.

To tune how close the random unitary is to the identity, we introduce a diagonal operator $\mathcal{D}$, with normally distributed real elements, such that the final noise term is
\begin{equation}
    U_{\mathcal{E}(\gamma)} = \exp(i \mathcal{D}\gamma )\mathcal{R} \ ,
\end{equation}
where $\gamma$ is a parameter that controls the magnitude of the error. The result is a dense unitary random matrix, whose eigenvalues lie on the unitary circumference in the complex plane. One of them is $-1$, originating from the Householder reflection. 
The others are centered around $1$ with a width depending on $\gamma$.
We characterize the average difference between the $U_{\mathcal{E}(\gamma)}$ and the identity through the Hilbert--Schmidt inner product
\begin{eqnarray}\label{eq:HS}
    d_{\rm HS}(\gamma)=\frac{\overline{\langle U_{\mathcal{E}(\gamma)},\mathbb{1}\rangle_\mathrm{HS}}}{{\rm dim}\mathcal{H}} = \left(1-\frac{2}{{\rm dim}\mathcal{H}}\right) \nep^{-\gamma^2/2}\ .
\end{eqnarray}

Using this ensemble of random unitaries, the noisy discrete-time evolution over a single Trotter step reads
\begin{equation}\label{eq:time_evo}
    U(\Delta t) = U_{\mathcal{E}(\gamma)}\nep^{-i H_E \Delta t}\nep^{-i H_B \Delta t}\ ,
\end{equation}
with $U_{\mathcal{E}(\gamma)}$ being drawn independently for every Trotter step of every trajectory.
The ensemble average of Eq.~\eqref{eq:time_evo} constitutes the error channel used in Eq.~\eqref{eq:gen_Ham}.

A realistic estimate for the error rate strongly depends on the individual qudit hardware. 
Reference~\cite{GonzalezCuadra_PRL2022} also consider a small lattice with a pure gauge theory and a discrete group, where the main sources of error are control-multiplications used to implement the plaquette operators. On a Rydberg-atom qudit simulator, they estimate the achievable fidelity to be $99.6\%$. 
Since six of such two-qudit gates are required for each plaquette operator, one can expect the overall fidelity of a single Trotter step to be about $95\%$ for the two-plaquette lattice we have considered in this work.
This corresponds to a noise strength $\gamma \sim 0.22$, which is compatible with the regime in which both PPS and DPS are found to work. At the same time, the bare noisy dynamics quickly relaxes to a nonphysical steady state, highlighting the necessity of error-mitigation schemes.

\section{Gauge protection on three plaquettes}\label{app:3p}
Any error mitigation technique is expected to become more resource-expensive for increasing system size, as the total error rate typically increases.
In this appendix, we benchmark DPS and PSV for a larger system composed of a ladder with three plaquettes and periodic boundary conditions, consisting of 6 qudits in total, and a Hilbert space of dimension 46656.
To reduce the memory requirements, we consider a global dephasing noise instead of dense random unitaries.
We model the dephasing noise through a Hamiltonian
$H_{\textrm{dp}}$ that is diagonal in the group element basis and whose entries are sampled from a normal distribution with vanishing mean and unit variance.
$H_{\textrm{dp}}$ is independently drawn at each Trotter step.
The noisy discrete-time evolution for a single step then reads
\begin{equation}\label{eq:time_evo_deph}
    U(\Delta t) = \nep^{-i H_{\rm dp}\gamma}\nep^{-i H_E \Delta t}\nep^{-i H_B \Delta t}\ ,
\end{equation}

First, we study the effect of dephasing noise on the target dynamics and the level of gauge protection offered by the two techniques in the two-plaquette ladder investigated in the main text. 
We show the ensemble average over $M=5000$ trajectories in Fig.~\ref{fig:3_plaquettes}(a), with noise strength $\gamma=0.2$ and the same parameters used throughout this paper ($1/g^2=0.5$ and $\Delta t/a=0.25)$. 
Despite the simpler structure than the dense random unitaries considered in the main text, the noisy dynamics is qualitatively similar. DPS and PSV both yield excellent error mitigation, with the former performing slightly better at longer times, similar to Fig.~\ref{fig:cover_pic}(b).

The overall picture continues to hold when considering the larger system with 3 plaquettes, as reported in Fig.~\ref{fig:3_plaquettes}(b), where we again monitor the left plaquette of the lattice.
In DPS, we again measure only one local symmetry after each Trotter step so that the overall cost of the mitigation technique does not increase with respect to the two-plaquette model. 
Remarkably, the reduced measurement density only slightly affects the gauge protection, compared to the smaller system shown in Fig.~\ref{fig:3_plaquettes}(a). 
For a fair comparison, the noise strength here is rescaled such that the overall error rate $\sim \gamma^2$ is proportional to the system size. 
Given a small local error probability $p_{\rm err}$, the error rate per Trotter step for $N$ qudits at first order in $p_{\rm err}$ is
given by $\gamma^2\simeq 1-(1-p_{\rm err})^N\simeq N p_{\rm err}$.
Here, $p_{\rm err}$ reflects the average fault rate per qudit resulting from all gates applied in the implementation of a single Trotter step. When increasing the system size from two to three plaquettes, we assume that this microscopic error rate $p_{\rm err}$ remains fixed. With the number of operations per layer growing linearly with the number of plaquettes, this leads to a corresponding linear increase of the overall infidelity $\gamma^2$ for each Trotter step, i.e.
$\gamma^2_{3\mathrm{P}} = 3\gamma^2_{2{\rm P}}/2$.
Because the measurement density is also rescaled down with the same ratio $3/2$, the decay of the DPS survival probability behaves similarly in the two cases, as shown in the right panel of Fig.~\ref{fig:3_plaquettes}(c).

Notably, PSV now better approximates the exact time evolution with respect to the data shown for the two-plaquette lattice. 
This counterintuitive behaviour can be explained by considering the ratio between the dimensions of the physical sector and the total Hilbert space. 
In the two-plaquette lattice, one has $\frac{\textrm{dim}\mathcal{H}_{\textrm{P}}}{\textrm{dim}\mathcal{H}_{\textrm{tot}}} = \frac{49}{1296} \simeq 0.0378$, while in the three-plaquette system $\frac{\textrm{dim}\mathcal{H}_{\textrm{P}}}{\textrm{dim}\mathcal{H}_{\textrm{tot}}} = \frac{251}{46656} \simeq 0.0054$. 
The smaller fraction reduces the probability of effective gauge-invariant errors originating from two consecutive gauge-violating errors that cancel, which PSV is not able to reveal. Hence, the fraction of noisy states projected into the physical sector retains information about the physical dynamics for a longer time.
However, considering the overhead scaling with the number of vertices discussed in Sec.~\ref{sec:scaling} suggests DPS could hold a practical advantage if mid-circuit measurements are available with sufficient speed and accuracy.
Interestingly, the weight of the symmetrized density matrix $\tr[\Pi_\s\rho]$ also does not display appreciable differences between the two- and three-plaquette ladders, see left panel of Fig.~\ref{fig:3_plaquettes}(c).

\bibliography{references}

\end{document}